# Compositional Complexity-Induced Ultralow Friction in Medium-Entropy MXenes

Jiaoli Li [1], Yuwei Zhang [2], Congjie Wei [1], Yanxiao Li [3], Shuo He [4], Risheng Wang [2], Brian Wyatt [5], Reza Namakian [6], Babak Anasori [5,7], Kelvin Xie [8], Tobin Filleter [4], Ali Erdemir [6], Wei Gao [6], Chenglin Wu [*,1]

[1]Zachry Department of Civil and Environmental Engineering, Texas A&M University, College Station, TX, 77843, USA

[2]Department of Chemistry, Missouri University of Science and Technology, Rolla, MO, 65401, USA

[3]National Energy Technology Laboratory (NETL), NETL Support Contractor, Pittsburgh, PA, USA

[4]Department of Mechanical & Industrial Engineering, University of Toronto, Toronto M5S 3G8, CA

[5]School of Materials Engineering, Purdue University, West Lafayette, IN, 47907, USA

[6]J. Mike Walker '66 Department of Mechanical Engineering, Texas A&M University, College Station, TX 77843, USA

[7]School of Mechanical Engineering, Purdue University, West Lafayette, IN, 47907, USA

[8]Department of Materials Science and Engineering, Texas A&M University, College Station, TX, 77843, USA

[*]Corresponding author E-mail: chenglinwu@tamu.edu

**Abstract**

Two-dimensional MXenes are promising solid lubricants, yet the roles of compositional complexity and surface chemistry in governing their interfacial friction remain poorly understood. Here, we systematically investigate the adhesion and friction behaviors of medium-entropy (ME) MXenes (TiVNbMoC$_3$ and TiVCrMoC$_3$) and compare them with conventional titanium carbide MXenes (Ti$_2$C and Ti$_3$C$_2$) using a SiO$_2$ colloidal atomic force microscopy probe. Thermal annealing at 200 °C induces a conversion of surface terminations from –OH to –O groups, leading to a pronounced reduction in adhesion energy and friction force across all MXenes studied. ME MXenes exhibit larger adhesion reductions due to their higher initial –OH contents and more extensive –OH-to-–O conversion. In addition, the intrinsically higher out-of-plane bending stiffness of ME MXenes further suppresses energy dissipation during sliding, promoting ultralow friction. Remarkably, superlubricity is achieved for the first time in ME MXenes, with annealed TiVCrMoC$_3$ exhibiting a coefficient of friction as low as 0.0022—surpassing graphene, MoSe$_2$, and other MXenes evaluated using the same experimental approach. These findings demonstrate that compositional complexity provides a powerful design strategy for engineering MXenes with exceptional tribological performance and establish ME MXenes as a new class of solid lubricants.

## INTRODUCTION

Compositionally complex MXenes, including medium and high-entropy MXenes (ME- and HE-MXenes)[1], share a similar concept with the high-entropy alloys by incorporating four or more transition metals in near-equiatomic proportions[2,3]. This unique composition yields a homogenous, single-phase structure with a stable crystal lattice and a diverse array of terminating groups. The resulting compositionally complex MXenes are anticipated to exhibit unique physical and chemical characteristics, making them suitable for various applications where mechanical strength, stability, and tribological properties are essential[4].

Compared with well-known single or double-transition metal carbide MXenes, like $Ti_2CT_x$, $Ti_3C_2T_x$, and $Mo_2TiC_2T_x$, the compositionally complex MXenes can have relatively large lattice parameter variations and higher crystal disorder due to their atomistic mismatches among the incorporated transition metals. Their increased crystal disorder, with rich surface terminating groups, can lead to excellent mechanical and tribological behaviors[5-7] for potential applications as solid lubricants and self-lubricating composite materials[8,9]. The first and foremost question is about their surface adhesion and friction behavior as compared to Ti-C MXenes.

For Ti-C MXenes, both out-of-plane bending stiffness and surface-terminating groups (i.e., -OH and -O groups) determine the surface adhesion and friction. In comparison with graphene and $MoSe_2$, Ti-C MXenes have relatively larger out-of-plane bending stiffness, resulting in less number-of-layer dependency[10-14]. The -OH surface-terminating groups on Ti-C MXenes increase the adhesion and friction due to hydrogen bonding and water bridging[12,15-18]. Meanwhile, -O surface-terminating groups decrease the adhesion and friction[19]. This has led researchers to explore surface-terminating group alteration through thermal annealing to tune the tribological performance of MXenes[20-23].

For the compositionally complex (i.e., ME and HE) MXenes[1], HF etching is typically expected, which could lead to a high concentration of -OH terminating groups[24-26]. This high concentration could lead to enhanced surface adhesion and friction due to hydrogen bonding. The multi-metallic element composition, high-entropy metal-carbide binding, and the increased monolayer thickness could lead to the high out-of-plane stiffness, in turn, reduce the puckering effect[10,27,28], and lower friction. These two tuning knobs could lead to complex tribological behaviors of the ME and HE MXenes compared with Ti-C MXenes. However, no experimental results are currently available to elucidate this complexity.

To reveal the fundamental nanoscale tribological behavior of compositionally more complex (i.e., ME and HE) MXenes, we conducted the first investigations on the structure and surface-to-tribology relationship of the ME MXenes, specifically $TiVNbMoC_3$, and $TiVCrMoC_3$, and compared them to the $Ti_2C$ and $Ti_3C_2$ MXenes. The atomic arrangement and crystal structures of the ME MXenes were characterized by the High-Angle Annular Dark-Field Scanning Transmission Electron Microscopy (HAADF-STEM) and X-ray Diffraction (XRD). The surface morphology, layer thickness measurements, layer count, and detailed assessment of surface roughness were characterized by the atomic force microscope (AFM). The X-ray Photoelectron Spectroscopy (XPS) results provide quantifiable information on surface terminations (-OH, -O, and -F) through peak deconvolution and intensity ratios. The effect of surface terminations on tribological performance was investigated through surface-annealing-induced conversion of terminating groups. At an annealing temperature of 200 °C, the -OH group was effectively reduced by 71-91%, as verified by XPS. The adhesion and friction of ME MXenes and Ti-C MXenes were characterized by AFM with colloidal probes, following our previous approach[12-14]. After annealing, ME MXenes demonstrated a significant reduction in adhesion and friction. Notably, annealed TiVCrMoC$_3$ exhibited the lowest coefficient of friction (0.0022) among all the MXenes investigated, demonstrating superlubricity and highlighting its potential as an exceptional solid lubricant.

## RESULTS

*Materials characterization*

**HAADF-STEM atomic structure characterization:** The atomic structure models, abTEM-simulated HAADF micrographs, and experimental HAADF micrographs for TiVNbMoC$_3$, TiVCrMoC$_3$, Ti$_2$C, and Ti$_3$C$_2$ MXenes at fresh conditions are presented in **Figure 1**. The abTEM simulations[29] recapitulate the crystal symmetry and lattice periodicity and show statistical agreement with the contrast distributions in experimental HAADF-STEM micrographs, consistent with random site occupancy by transition metal species of differing atomic number (Z). In TiVNbMoC$_3$ and TiVCrMoC$_3$, the four transition metals (Ti, V, Nb/Cr, Mo) are randomly distributed, as illustrated in **Figure 1(a1)** and **1(b1)**. The in-plane structure exhibits hexagonal symmetry, like the MAX-phase precursors (**Figure 1(a2)** and **1(b2)**). The simulated HAADF images (**Figure 1(a3) and 1(b3)**) using the abTEM simulator and experimentally acquired images (**Figure 1(a4) and 1(b4)**) highlight Z-contrast, revealing the in-plane locations of the metallic species (Ti, V, Nb/Cr, and Mo). Due to the smaller Z of carbon, its contribution to the HAADF signal is negligible. The non-uniform brightness observed in each atomic column results from the random distribution of various species of transition metal atoms, confirming the medium-entropy nature of the material.

In contrast, Ti-C MXenes have a simpler structure where each layer consists of one type of metallic atom (Ti in this study) and C atoms. These layers stack to form a two-dimensional material after removing the Al layer from their MAX phase. Specifically, $Ti_2C$ consists of one layer of Ti and one layer of C, as illustrated in **Figure 1(c1)**, while $Ti_3C_2$ has a structure where the C atoms are sandwiched between layers of Ti atoms, as shown in **Figure 1(d1)**. The crystal structures of the Ti-C MXenes are also hexagonal, as seen in **Figure 1(c2)** for $Ti_2C$ and **Figure 1(d2)** for $Ti_3C_2$. Similarly, the simulated (**Figure 1(c3)** for $Ti_2C$ and **Figure 1(d3)** for $Ti_3C_2$) and HAADF images (**Figure 1(c4)** for $Ti_2C$ and **Figure 1(d4)** for $Ti_3C_2$) of the Ti-C MXenes clearly show bright Ti atom columns with a uniform brightness due to the higher atomic number of titanium compared to carbon. From the HAADF images analysis, the lattice parameters of $TiVNbMoC_3$, determined through fast Fourier transformation (FFT) analysis, were found to be an a-lattice parameter (a-LP) of 3.036 Å. Similarly, for $TiVCrMoC_3$, the measured lattice parameter was a-LP = 2.97 Å. These values are comparable to $Ti_2C$ (a-LP: 3.08 Å) and $Ti_3C_2$ (a-LP: 3.07 Å), highlighting the in-plane structural similarities between the ME MXenes and Ti-C MXenes.

**XRD atomic layer characterization.** The synthesis of MXenes involves etching the Al layer from their corresponding MAX phases, as illustrated in **Figure 1e** for ME MXenes and Ti-C MXenes. The XRD spectra for different types of MXenes and their respective MAX phases are shown in **Figure 1f** to **1i**, confirming the successful removal of Al elements. Specifically, **Figure 1f** and **1g** present the XRD patterns for $TiVNbMoC_3$ and $TiVCrMoC_3$, where the main (002) peak exhibits a noticeable left shift, indicating the selective etching of Al atoms from the MAX phases. Additionally, the (104) peak is another distinguishing feature between ME MXenes and their MAX phases, with the corresponding peaks in the ME MXenes being weaker due to reduced Al content. Based on Bragg's law[30], the (002) peaks of $TiVNbMoC_3$ and $TiVCrMoC_3$ correspond to d-spacings of 22.65 Å and 22.22 Å, respectively.

Similarly, the (002) peaks of $Ti_2C$ (**Figure 1h**) and $Ti_3C_2$ (**Figure 1i**) shift from 9.63° to 6.64° and from 13° to 6.19°, respectively, due to the selective removal of Al atoms from $Ti_2AlC$ and $Ti_3AlC_2$. The weakening of the (104) peaks in $Ti_2C$ and $Ti_3C_2$ MXenes further indicates the depletion of their MAX phase precursors, $Ti_2AlC$ and $Ti_3AlC_2$[31]. In addition to distinguishing MXenes from their MAX phases, the d-spacing of the MXene sheets can be determined from the (002) peak in the XRD spectrum. For $Ti_2C$ and $Ti_3C_2$, the calculated d-spacings are 13.3 Å and 14.26 Å, respectively. The layered structures of both ME and Ti-C MXenes result from removing Al from their MAX phases. However, due to an increased number of layers of transition metals, the d-spacing in ME MXenes is noticeably larger than in Ti-C MXenes.

**AFM morphology characterization.** AFM was performed on the MXene to characterize its morphology, layer thickness, and surface roughness. **Figure 2(a1) and 2(b1)** displays the stacked TiVNbMoC$_3$ and TiVCrMoC$_3$ MXene flakes with various monolayers (calculated by thickness over d-spacing). The average flake size observed was about 5 $\mu$m, with the monolayer thickness approximately 2.32 nm for TiVNbMoC$_3$ (**Figure 2(a2)**, and 2.28 nm for TiVCrMoC$_3$ (**Figure 2(b2)**). The surface roughness of the ME MXene flakes remains around 70 pm for layers 1-5, with a slight decrease. When the number of layers increases to 10, the surface roughness decreases to about 57 pm, as shown in **Figure 2(a3)** for TiVNbMoC$_3$ and **Figure 2(b3)** for TiVCrMoC$_3$. This reduction in surface roughness with increasing thickness can be attributed to the diminishing influence of the substrate surface effects[32].

In contrast, the monolayer thickness of Ti-C MXenes is measured at around 1.35 nm for Ti$_2$C MXene (**Figure 2(c2)**) and 1.42 nm for Ti$_3$C$_2$ MXene (**Figure 2(d2)**). The monolayer thickness of Ti-C MXenes is thinner than that of the ME MXenes. The surface roughness of Ti-C MXenes is around 80 pm for layers between 1 and 5 (**Figure 2(c3)** for Ti$_2$C and **Figure 2(d3)** for Ti$_3$C$_2$), which is larger than that of ME MXenes. The flake size of Ti-C MXenes is approximately 500 nm (**Figure 2(c1)** and **2(d1)**), which is smaller than that of ME MXenes. This difference in flake size could be due to the different synthesis methods, as Ti-C MXenes were synthesized using a mixture of HCl and LiF, whereas ME MXenes were synthesized using direct HF etching.

The 200 °C annealing was conducted on all MXenes. The morphology of the annealed MXene nanosheets is shown in **Figure S2**. The annealing process preserves their intrinsic sheet-like morphology and topographical features. The MXene layer thickness increased slightly after annealing. The thickness of ME MXene increased by approximately 0.6 nm, from 2.32 nm to 2.98 nm for TiVNbMoC$_3$ and from 2.28 nm to 2.89 nm for TiVCrMoC$_3$. Similarly, the layer thickness of Ti-C MXenes also increased (around 1 nm), from 1.35 nm to 2.28 nm for Ti$_2$C MXene and from 1.42 nm to 2.49 nm for Ti$_3$C$_2$ MXene. While annealing usually reduces layer thickness by removing hydrogen-containing surface terminations, the observed increase is likely due to altered MXene-substrate interactions. The loss of hydrogen bonding may weaken adhesion, contributing to the apparent increase in thickness. This indicates that annealing affects both termination chemistry and substrate interaction.

**XPS surface terminating groups characterization.** High-resolution XPS was conducted to analyze the chemical bonding of fresh and annealed MXenes.

Fresh conditions: For ME MXenes at fresh conditions, survey spectra in **Figure S3** (TiVNbMoC$_3$) and **S4** (TiVCrMoC$_3$) revealed the presence of elements such as F 1s, Nb 3d/Cr 2p, O 1s, V 2p, Ti 2p, C 1s, and Mo 3d. Deconvolution of the Ti 2p region confirmed the presence of Ti-C and TiO$_2$ bonds, while the V 2p spectrum showed peaks corresponding to V$_2$O$_5$, V$^{2+}$, and V$^{4+}$ [33]. In the Nb 3d spectrum for TiVNbMoC$_3$ (**Figure S3d**), six chemical states were identified, including Nb, Nb(I/II/IV), NbO, Nb$^{3+}$-O, Nb$^{4+}$-O, and Nb$_2$O$_5$. In contrast, the Cr 2p spectrum for TiVCrMoC$_3$ (**Figure S4d**) revealed peaks for Cr-C and Cr-T$_x$. Peaks in the Mo 3d region (**Figure S3e** and **S4e**) were assigned to Mo$^{5+}$, Mo$^{6+}$, C-Mo-T$_x$, and Mo. The C 1s spectrum (**Figure S3f** and **S4f**) indicated C-Mo/Ti-T$_x$, C-C, CH$_x$, and C-O bonds, while the O 1s region (**Figure 3(b1)** and **3(c1)**) highlighted TiO$_2$/MO$_x$, C-M-O$_x$, and C-M-(OH)$_x$ groups. The F 1s region (**Figure 3(b2)** and **3(c2)**) revealed M-F and Al-F bonds. The minor Al-F peak indicates small amount of surface residual due to HF etching.

For Ti-C MXenes at fresh conditions, the XPS survey spectra in **Figure S5** (Ti$_2$C) and **S6** (Ti$_3$C$_2$) identified F 1s, O 1s, Ti 2p, and C 1s as key elements. Deconvolution of the Ti 2p spectrum (**Figure S5b** and **S6b**) revealed Ti-C, Ti (II), Ti (III), and Ti-O states[34]. The C 1s spectrum (**Figure S5c** and **S6c**) confirmed the presence of C-C and C-Ti-T$_x$ bonds, while the O 1s spectrum (**Figure 3(d1)** and **3(e1)**) highlighted TiO$_2$, C-Ti-O$_x$, and C-Ti-OH groups[35]. In the F 1s spectrum (**Figure 3(d2)** and **3(e2)**), peaks were assigned to Ti-F and Al-F.

The XPS results demonstrate that the chemical composition and bonding environments differ significantly between ME MXenes and Ti-C MXenes. ME MXenes, due to their multiple transition metals, exhibit more complex oxidation states and bonding environments, such as Nb-O, V-O, and Mo-O, which are absent in Ti-C MXenes. These differences are further emphasized by their broader range of terminating groups and their relative distributions. As summarized in **Figure 3f**, fresh ME MXenes (TiVNbMoC$_3$ and TiVCrMoC$_3$) exhibited higher -OH group content (70% and 69%, respectively) compared to Ti$_2$C (61%) and Ti$_3$C$_2$ (41%), while Ti-C MXenes displayed higher -F group content (21% for Ti$_2$C and 33% for Ti$_3$C$_2$) than ME MXenes (15% for TiVNbMoC$_3$ and 21% for TiVCrMoC$_3$).

Annealed conditions: The XPS results of annealed MXenes are shown in **Figure S7** (TiVNbMoC$_3$), **S8** (TiVCrMoC$_3$), **S9** (Ti$_2$C), and **S10** (Ti$_3$C$_2$). The annealing transformed -OH groups into -O groups, as confirmed by deconvoluted XPS spectra. Despite these changes, the chemical states of core elements, including Ti, V, Nb, Cr, Mo, and C, remained largely unchanged, indicating that annealing primarily alters surface terminations while maintaining the structural integrity of the transition metal-carbon framework. The relative ratios of -OH, -O, and -F groups (**Figure 3b-3f**) revealed a substantial reduction in -OH groups and a corresponding increase in -O groups across all MXenes after annealing. The -O group became the dominant surface termination, accounting for 84% and 79% in Ti$_2$C and Ti$_3$C$_2$, respectively, and 78% in TiVNbMoC$_3$ and TiVCrMoC$_3$. The -F group content also decreased significantly post-annealing, with Ti-C MXenes showing reductions from 21% to 10% (Ti$_2$C) and 33% to 16% (Ti$_3$C$_2$). ME MXenes exhibited even lower -F content after annealing, with TiVNbMoC$_3$ decreasing from 15% to 5% and TiVCrMoC$_3$ from 21% to 1%.

*Adhesion of MXenes*

The AFM measured adhesion force and energy for all MXenes under fresh and annealed conditions are shown in **Figure 4a**, and **4c-4d**.

**Adhesion force.** The obtained force *versus* height measurement curves are shown in **Figure S11** (fresh MXenes) and **S12** (annealed MXenes). Each curve has six stages: approach, jump-to-contact, contact compression, peak force, adhesion pull-off, and detachment, as defined in our previous work[12]. The peak force ((V) in **Figure 4a**) observed during the withdrawal process gives the adhesion force. The tabulated adhesion forces for all MXenes measured are shown in **Table S1**.

The adhesion forces of fresh ME MXenes (36.43 nN ~ 43.74 nN for TiVNbMoC$_3$ and 49.76 nN ~ 53.68 nN for TiVCrMoC$_3$) are larger than those of Ti-C MXenes (15.71 nN ~ 25.12 nN for Ti$_2$C, and 30.56 nN ~ 37.10 nN for Ti$_3$C$_2$). No layer-dependent adhesion force was observed for any of the MXenes. The adhesion force of the annealed MXenes is significantly lower than that of the fresh ones. We observed 72, 67, 70, and 74% reductions for TiVNbMoC$_3$, TiVCrMoC$_3$, Ti$_2$C, and Ti$_3$C$_2$, respectively.

**Adhesion energy.** The Rumpf model[36,37] was adopted to calculate the adhesion energy, given that the surface roughness is much smaller than the tip radius (the maximum surface roughness is 72 and 81 pm for ME and Ti-C MXenes, respectively). The calculation details are shown in

**Supplementary Section 3**. The interaction ranges are measured as the distance between the jump-off point and zero-force recovery, as shown in **Figure 4a**.

Adhesion energy of fresh MXenes: The average adhesion energies *versus* interaction ranges are shown in **Figure 4c** and **4d**. For fresh MXenes, the average adhesion energies are 1.262, 1.232, 0.515, and 1.115 J/m², for TiVNbMoC$_3$, TiVCrMoC$_3$, Ti$_2$C, and Ti$_3$C$_2$, respectively. The corresponding interaction ranges are 8.27, 5.81, 7.09, and 5.62 nm. Ti$_2$C has the lowest adhesion energy, while TiVNbMoC$_3$ exhibits the highest adhesion energy and the largest interaction range of 8.27 nm. In contrast, Ti$_3$C$_2$ and TiVCrMoC$_3$ share a similar interaction range, the smallest among all MXenes. Unlike graphene, the adhesion energy does not vary significantly with the number of monolayers for ME MXenes and Ti-C MXenes[12].

Adhesion energy of annealed MXenes: The calculated adhesion energy of the annealed MXenes is shown in **Figure S12a** to **S12d** with the Gaussian fitting of the measurement histograms (shown in **Figure S12e-S12h**). The average adhesion energies of the annealed MXenes are 0.972, 0.935, 0.42, and 0.833 J/m² for TiVNbMoC$_3$, TiVCrMoC$_3$, Ti$_2$C, and Ti$_3$C$_2$, respectively. The corresponding interaction ranges are 8.17, 5.71, 5.91, and 5.45 nm. Compared to the fresh state, we observed reductions of 23, 24, 18, and 25% in adhesion energy for TiVNbMoC$_3$, TiVCrMoC$_3$, Ti$_2$C, and Ti$_3$C$_2$, respectively.

The adhesion energy for the annealed MXenes, as incorporated in **Figure 4c**, shows a clear decrease compared to the corresponding fresh MXenes. However, the ranking in adhesion energy remains after annealing. Like the fresh condition, annealed Ti$_2$C shows the lowest adhesion energy and interaction range, while annealed TiVNbMoC$_3$ shows the largest ones (shown in **Figure 4d**). Interestingly, while Ti$_2$C shows a slight reduction in interaction range, the other three MXenes maintain almost consistent interaction ranges after annealing. These results clearly indicate that annealing will lower surface adhesion for all MXenes. However, this factor alone does not warrant any friction reduction for ME MXenes since the annealed ME MXenes still show 34% higher adhesion energy, and 18% longer interaction ranges compared with Ti-C MXenes.

*Friction of MXenes*

**Friction force**. The friction performance of ME MXenes and Ti-C MXenes was characterized using AFM lateral measurements (**Figure 4b**) under varying normal forces and number-of-monolayer. Normal loads ranging from 0 $\mu$N to 1.5 $\mu$N were applied to evaluate the frictional response of both fresh and annealed MXene nanosheets across different numbers of monolayers. The resulting measurement curve comprises four distinct stages as illustrated in **Figure 4b** (left): an initial sharp increase as the AFM tip engages with the surface, a flat plateau during steady sliding, a sharp decrease as the tip disengages, and a final flat plateau as the tip fully retracts. The following analysis compares the obtained friction behavior of all MXenes under fresh and annealed conditions, focusing on the effects of layer numbers, applied normal forces, and terminating groups.

Friction force of fresh MXenes: The voltage *versus* scan distance curves are presented in **Figure S13-S16**. The increased gap between trace and retrace revealed that the friction force between MXene nanosheets and AFM tip keeps increasing with the normal force increases. In contrast, compared with the friction measurement of MXenes at one layer, the gap between trace and retrace in corresponding to 14-layer MXene showed a downward trend, demonstrating that with the number of layers increasing, friction force keeps decreasing. The obtained friction measurement loops of ME MXene (**Figure S13** for $TiVNbMoC_3$, and **S14** for $TiVCrMoC_3$) and Ti-C MXenes (**Figure S15** for $Ti_2C$, and **S16** for $Ti_3C_2$) reflected a similar increasing trend in the normal force direction and a decreasing trend in the number of layer directions. The friction force of 14-layer MXenes is summarized in **Figure 4e** (represented by circles). The linear fitting curves applied to these figures demonstrate a high coefficient of determination ($R^2$), indicating a strong linear relationship between the friction force and the applied normal force.

Friction force of annealed MXenes: The measured loops of the annealed MXenes with different layers are shown in **Figure S17** ($TiVNbMoC_3$), **S18** ($TiVCrMoC_3$), **S19** ($Ti_2C$), and **S20** ($Ti_3C_2$). For the same MXene, with the normal forces increasing, the gap between trace and retrace keeps increasing, and with the number of layers increasing, the gap between trace and retrace continuously decreases. All these trends in annealed MXenes are the same as those of fresh MXenes. The triangles and fitting curves in **Figure 4e** represent the calculated friction forces of annealed 14-layer MXenes, illustrating a linear relationship between the friction force and the applied normal force.

The average friction forces for fresh and annealed (in parentheses) MXenes with 14 layers are 132.37 (123.67), 127.58 (107.6), 107.41 (80.12), and 112.15 (94.02) nN, for $TiVNbMoC_3$, $TiVCrMoC_3$, $Ti_2C$, and $Ti_3C_2$, respectively. An average reduction of 11% and 21% was observed for ME MXenes and Ti-C MXenes, respectively.

**Coefficient of friction.** The CoF is defined as:

$$CoF = \frac{f_{lat} - f_{adh}}{f_{nor}} \quad (1)$$

Where $f_{lat}$ and $f_{nor}$ are the friction force obtained from lateral measurements and the corresponding applied normal force. $f_{adh}$ is the corresponding adhesion force. It should be noted that CoF is within the range of 0~1, and materials with CoF smaller than 0.01 are considered superlubricious. The friction measurements under 0.9 $\mu$N normal forces were adopted to calculate the CoF of all four MXenes with different numbers of layers under fresh and annealed conditions. The results are shown in **Figure 4f** (left). The tabulated CoFs for all MXenes measured are shown in **Table S2**. For fresh MXenes, the CoF decreases with increasing monolayer number, stabilizing at ~5 layers for both ME and Ti-C MXenes. In annealed MXenes, the CoF continues to decrease with increasing monolayer number until ~15 layers.

The CoFs of annealed MXenes show a pronounced reduction compared to their fresh counterparts, as presented in **Figure 4f** (right). Annealed TiVCrMoC$_3$ exhibits the lowest CoF. For all cases, the annealing process significantly reduces the CoFs of all MXenes, with ME MXenes showing a decrease from 0.0679 to 0.0140 for TiVNbMoC$_3$ and from 0.0585 to 0.0022 for TiVCrMoC$_3$, and Ti-C MXenes decreasing from 0.0474 to 0.0041 for Ti$_2$C and from 0.0478 to 0.0181 for Ti$_3$C$_2$. This annealing reduced friction indicates that the intrinsic atomistic structure of the monolayer has a more dominant effect on MXene friction.

*Atomistic Modeling*

Both surface adhesion energy and out-of-plane deformations, such as bending and buckling, often influence the friction of 2D materials. We conducted atomistic modeling using both density functional theory (DFT) and molecular dynamics (MD) with machine-learning-trained interatomic potentials to understand differences in surface adhesion and bending stiffness among the investigated MXenes.

For surface adhesion, we performed MD calculations (**Figure 5a-5e** details are provided in **Supplementary Section 4.1**). The surface terminating group ratios obtained from the XPS experiments were adopted in the modeling. We observed a similar reduction in adhesion energy and associated hydrogen bond de-association at equilibrium for the annealed MXenes, as shown in **Figure 5e**. The ranking for the calculated interfacial adhesion energy for the annealed MXenes against hydrophilic $SiO_2$ energy follows $TiVCrMoC_3 \approx TiVNbMoC_3 > Ti_3C_2 > Ti_2C$, which shared a similar trend as measured experimentally ($TiVNbMoC_3 \approx TiVCrMoC_3 > Ti_3C_2 > Ti_2C$). The calculated adhesion energy values differ slightly from the measured values (**Figure 5e**), primarily due to water bridging arising from the relatively high moisture content (~50%) during the experiments.

The Young's modulus and monolayer thickness are also calculated in the DFT (details can be found in **Supplementary Section 4.2**) as shown in **Figure 5f-5h**. Considering the differences in monolayer thicknesses ($TiVCrMoC_3 \approx TiVNbMoC_3 > Ti_3C_2 > Ti_2C$) and the reduced adhesion energy due to the surface annealing, one might expect a sharp decreasing trend in CoFs. However, the material defects in terms of atomic vacancies have been observed in ME MXenes as shown in the TEM images in **Figure S1**. These defects lead to a significant reduction of Young's modulus of ME MXenes and partially reduce the puckering effect (plot in **Figure 5h**).

In sum, the adhesion energy and stiffness calculation results support the experimental finding; that is, the reduced adhesion energy and increased bending stiffness give the annealed $TiVCrMoC_3$ the lowest friction behavior.

*Friction of Fresh MXene with Steel Substrates-Ongoing work*

While the previous sections focused on interactions between MXenes and oxide-based probes (e.g., $SiO_2$), metallic interfaces are particularly relevant for real-world tribological applications, especially in environments where direct contact with metal surfaces occurs. Recent computational and experimental studies have highlighted how various metals interact with two-dimensional (2D) materials, significantly altering interfacial friction and adhesion behaviors[38-41]. To extend our understanding of MXene tribology in such contexts, we investigated friction between multi-layer ME MXene films and 440C stainless steel. This high-carbon martensitic steel is commonly used in aerospace components, including precision ball bearings[42,43]. To better approximate application conditions, ongoing work extends steel-MXene measurements from the nano to the macro scale; preliminary methods and qualitative observations are provided in **Supplementary Section 5**.

## DISCUSSION AND FUTURE WORK

We summarized existing investigations into the friction performance of different 2D materials[14] characterized using the identical approach described in this work as shown in **Figure 5i**, focusing on two key aspects: friction between $SiO_2$ and 2D materials, including graphene, $MoSe_2$, $Ti_2C$, $Ti_3C_2$, $TiVNbMoC_3$, and $TiVCrMoC_3$, and the friction between 2D materials, including Ti-C MXenes, graphene, and $MoSe_2$. A few discussions are offered here: (i) Compared to other 2D materials, fresh ME MXenes exhibit higher CoFs when interacting with $SiO_2$. This behavior is primarily attributed to the dominant influence of -OH groups on MXene surfaces, which enhance interfacial interactions and friction forces. (ii) Annealing significantly improves the friction performance of MXenes by transforming -OH groups into -O groups, thereby reducing interfacial shear. Annealed ME MXenes ($TiVCrMoC_3$) exhibit exceptional performance among tested MXenes, surpassing graphene and $MoSe_2$. (iii) The high intrinsic bending stiffness and the relatively low adhesion energy of ME MXenes lead to extremely low friction when annealed. Both the DFT calculation and experimental observations indicate that the low CoFs of ME MXenes after annealing support this finding. (iv) It is suspected the ME MXenes will show even lower CoFs when interacting with other 2D materials or themselves due to the similar findings observed for other 2D materials as shown in **Figure 5i** (right parts: friction between $Ti_2C/Ti_3C_2$ and 2D materials). Inter 2D materials friction remains an interesting yet unknown field for future investigations of ME MXenes and other compositionally complex MXenes. (v) Our extended study with 440C steel probes further reveals that friction at metal-MXene interfaces can vary significantly depending on surface chemistry and roughness. Future work will focus on mechanistic studies at metal interfaces to guide MXene-based lubricant design.

## CONCLUSION

We observed a major terminating group conversion and induced reduction in adhesion and friction of the entropy-stabilized ME $TiVNbMoC_3$ and $TiVCrMoC_3$ MXenes. (i) Overall, fresh ME MXenes exhibit higher -OH content than Ti-C MXenes (~69% vs. 41~61%) due to their complex multi-metal bonding environments. Upon annealing, -OH groups are effectively converted into dominant -O terminations (>78%), while -F content is significantly reduced, and core metal-oxygen bonds are preserved. (ii) Fresh ME MXenes exhibit higher adhesion energy and longer interaction ranges than Ti-C MXenes due to their higher -OH content, which enhances interfacial bonding. Annealing reduces the adhesion energy of both ME and Ti-C MXenes by 24% on average. (iii) Fresh ME MXenes exhibit higher friction forces and CoFs than Ti-C MXenes, despite their greater bending stiffness. Upon annealing, the conversion of -OH to -O reduces surface contributions, allowing bending stiffness to dominate and markedly lowering friction, achieving super-lubricity in TiVCrMoC$_3$ (CoF=0.0022). This makes the annealed ME MXenes the best solid lubricant candidate among other 2D materials tested using the same approach, including graphene, MoSe$_2$, Ti$_2$C, and Ti$_3$C$_2$.

## Materials and Methods

**Chemicals.** MAX phases (TiVNbMoAlC$_3$ with 400 mesh, TiVCrMoAlC$_3$ with 400 mesh, Ti$_2$AlC with 200 mesh, and Ti$_3$AlC$_2$ with 200 mesh) were purchased from Jilin 11 Technology Co., Ltd. (Changchun, Jilin, China). 48% hydrofluoric acid (695068), lithium fluoride (62497), and 25 wt. % tetramethylammonium hydroxide (TMAOH) solution (331635) was purchased from Sigma-Aldrich (St. Louis, MO, USA). Hydrochloric acid (A142-212) was purchased from Fisher Scientific (Pittsburgh, PA, USA).

**ME MXenes preparation.** TiVNbMoC$_3$ and TiVCrMoC$_3$ shared an identical preparation method. The synthesis process is based on a top-down strategy, which means selective etching away of Al atoms from the MAX phases (TiVNbMoAlC$_3$ and TiVCrMoAlC$_3$). The synthesis process starts with slowly adding 0.5 g of MAX to a polyethylene jar containing 10 mL 48% HF solution to ensure full mixing and avoid spattering. Then, keep the mixture at 55 °C with continuous stirring at 400 rpm for 4 days to support the etching reaction. Repeat washing the obtained reactant, followed by 6 mins centrifugation at 3500 rpm five times until the pH value reaches 6.0 to eliminate HF and unreacted MAX. The obtained sediments were used for flake intercalation. 5 wt. % TMAOH was added to the washed sediments and keep the mixture continuously stirring at 500 ~ 600 rpm for 24 h at 55 °C to do intercalation. After intercalation, keep washing the mixture five times (~250 mL with DI water) in a centrifuge at 3500 rpm for 6 mins until the pH value attains 6.0 to remove the TMAOH. The collected supernatant was used for sample preparation.

**Ti-C MXenes preparation.** The general process of Ti-C MXene is similar to that of ME MXenes, however, a LiF/HCl approach was employed to use Li as the delamination agent, instead of the HF + TMAOH method used for ME MXenes. Specifically, 0.8 g LiF was added to 10 mL 9 mol HCl and kept the solution stirring for 30 min to prepare HF etchant. Then, 0.5 g MAX phase of Ti-C MXenes (Ti$_2$AlC and Ti$_3$AlC$_2$) was added slowly to the mixture of HCl and LiF and leave it at 35 °C with continuous stirring at 400 rpm for 30 h to achieve Al etching and MXene intercalation.

After that, keep washing and centrifuging the mixture at 3500 rpm for 5 min until the pH value comes to 6.0 to remove the etchant. The obtained supernatant was adopted for sample preparation.

**Interfacial deposition method.** The obtained MXene solution was coated on the mica surface using the interfacial deposition method. 50 μL of fresh MXene colloidal solution was first added to the mixture of 50 mL DI water and 3 mL toluene (244511, Sigma-Aldrich, USA), keeping them continuously stirring at 450 rpm for 15 min. Then, slowly pour the mixture of MXene, DI water, and toluene into a baker that is filled with 400 mL DI water and already has a mica substrate placed at the bottom, and keep it standing still for 20 min. The MXene films could self-assemble at the interface between bottom water and top toluene due to the different regent densities. After that, slowly lift the mica substrate at the bottom of the baker and pass through the interfacial layer to evenly catch the self-assembled MXene film. Finally, the coated MXene samples were dried with argon floating to avoid oxidation.

**Annealing**. The annealing process was conducted in an inert environment to remove the -OH group to avoid unwanted chemical reactions. In this work, the annealing treatment for all four MXene flakes was conducted at 200 °C for 2h within constant argon flow to reduce hydroxyl termination while maintaining -O and -F groups.

**TEM imaging.** Transmission electron microscopy (JEOL-ARM 200) was used to characterize the morphology and nanostructures of MXenes. The acceleration voltage is 200 kV, and the point-to-point resolution is 0.38 nm.

**XRD detection**. XRD was performed to explore the lattice structure of the MXenes and their MAX phases using a Bruker D2-phaser X-ray diffractometer with a Cu K ($\alpha$) radiation wavelength of 1.541Å. The scans were carried out from 5° to 90° (2θ) using a step size of 0.0026° (2θ) with a time step of 137 s/step.

**AFM imaging.** Atomic force microscopy (Dimension Icon AFM, Bruker, USA) was conducted to characterize the topography of the sample using tapping mode. In tapping mode for AFM imaging, an AFM tip with a 10 nm radius (FORTA, AppNANO, USA) was adopted to scan the sample surface under 1 Hz. The surface roughness is the square root of the average of the squares of the deviations of the surface profile from the mean of the surface height[44].

**XPS detection.** High-resolution XPS measurements were carried out on a Kratos Axis 165 Photoelectron Spectrometer using an Mg X-ray source and standard spot size.

**Adhesion measurements.** For adhesion measurement, we adopted a spherical AFM tip and considered it as a conical indenter. Indentation experiments were applied to the surface of MXenes with a different number of monolayers *via* an AFM tip. With the spherical AFM probes moving in the normal direction, adhesion forces are reflected due to the bending of the tip cantilever. The force *versus* displacement response for tip approach and withdrawal measured on MXene samples with different layers was shown in **Figure S11** and **S12**, which were obtained at 500 nm Z-ramp size with 0.1 Hz ramp rate. Note that the Z changes in the figure represent only a portion of the 500 nm range, as the force-displacement curves were cropped to highlight the relevant section of the response for clarity. The adhesion performance measurements of each sample were selected from six different testing positions to obtain the average results. Adhesion forces between few-layer MXene films and steel were also quantified using AFM. 440C stainless steel beads with diameters ranging from 13 to 15 $\mu$m were affixed to the ends of tipless AFM cantilevers using a precision adhesive technique. Measurements were performed at a scan rate of 0.99 Hz and a constant approach/retract velocity of 1.98 $\mu$m/s, with no dwell time applied at the point of contact.

**Friction measurements.** Friction measurements were conducted using a 5 $\mu$m diameter spherical shape tip (ACTG-SiO2-A-5, AppNANO, USA) at 1 Hz for ME MXenes and 0.8 Hz for Ti-C MXenes, and the tip probe had a 90° rotation with respect to the sample surface during the friction process due to the high sensitivity performance. The collected information on the voltage *versus* scan distance was used for friction analysis. Specifically, the scan distance was set to 100 nm. The friction force between the AFM tip and MXene nanosheets with 1, 2, 3, 4, 5, and more than 10 layers was measured under a series of normal forces (0 $\mu$N, 0.3 $\mu$N, 0.6 $\mu$N, 0.9 $\mu$N, 1.2 $\mu$N, and 1.5 $\mu$N). During the measurement processes, the friction force can be reflected by the twist of the AFM cantilever, which will lead to changes in voltage output. The voltage *versus* scan distance relationship during the trace and retrace processes was recorded. The gap ($V_{trace} - V_{retrace}$) between the trace and retrace process was used for the friction force calculation to eliminate the effect of vertical offset in the friction loop[45,46]. Additionally, to evaluate the frictional interaction between few-layer MXene films and steel, the same steel-beaded AFM cantilevers were used following calibration. The applied normal force was systematically varied from 0 to 160 nN. All measurements were conducted under ambient conditions, and lateral force signals were recorded to quantify frictional responses.

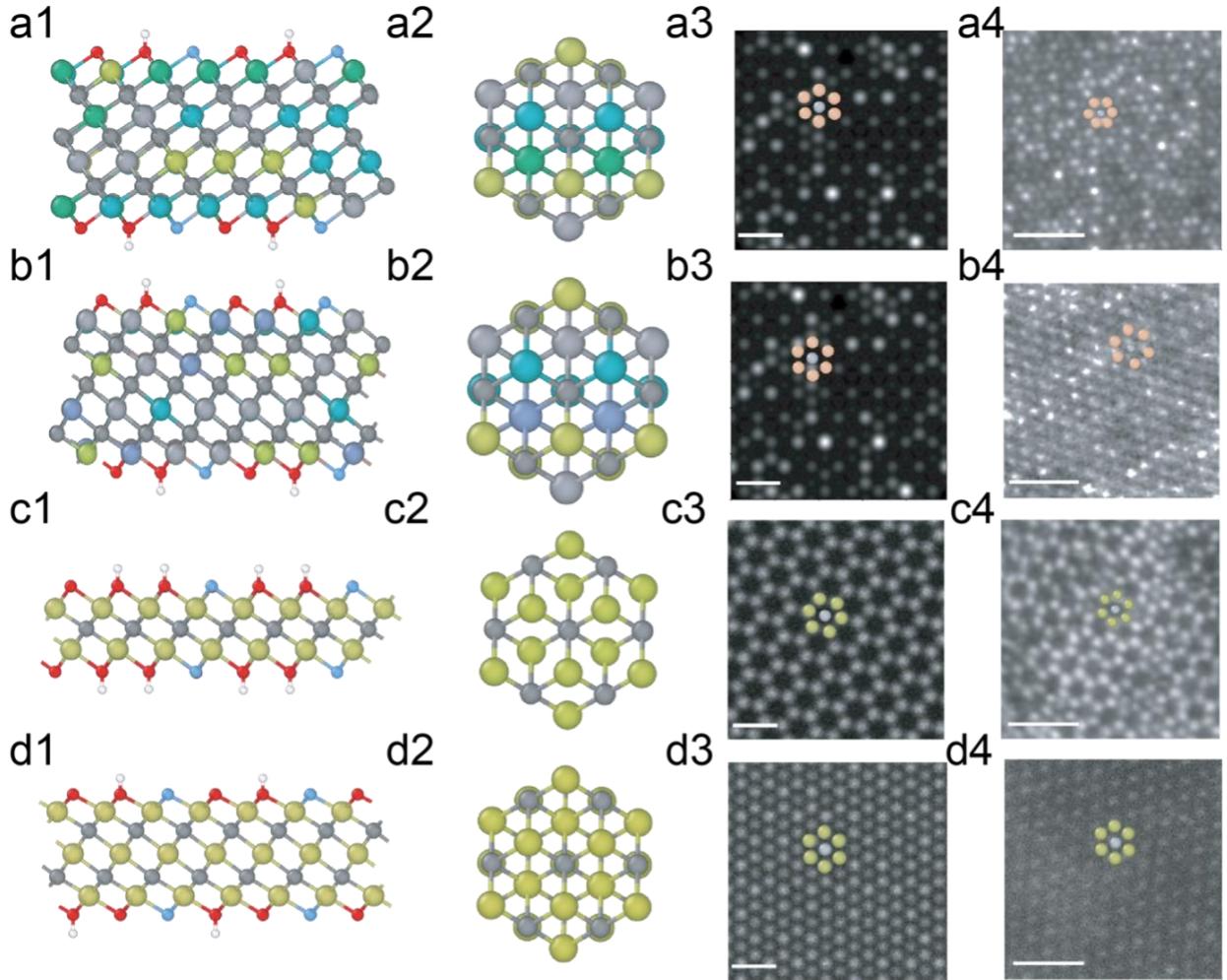

**Figure 1. Structural and chemical characterization of MXenes.** Atomic Structures of **a** TiVNbMoC$_3$, **b** TiVCrMoC$_3$, **c** Ti$_2$C, and **d** Ti$_3$C$_2$ MXene. For each MXene, the panels from left to right show cross-sectional atomic structure (1), in-plane atomic structure (2), simulated HAADF images (3), and experimentally acquired HAADF images (4). The scale bar is 1 nm. Note the terminating groups are also shown in the cross-sectional atomic structure models; **e** Schematic of the atomic structures of MXenes and their MAX phase, illustrating the etching process; XRD spectrum for MXenes and their MAX phases: **f** TiVNbMoC$_3$, **g** TiVCrMoC$_3$, **h** Ti$_2$C, and **i** Ti$_3$C$_2$ MXene.

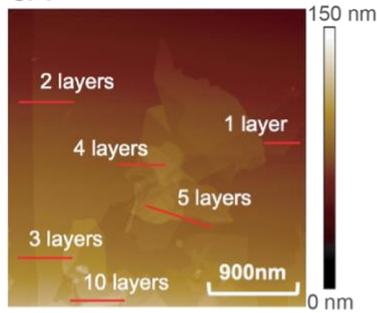 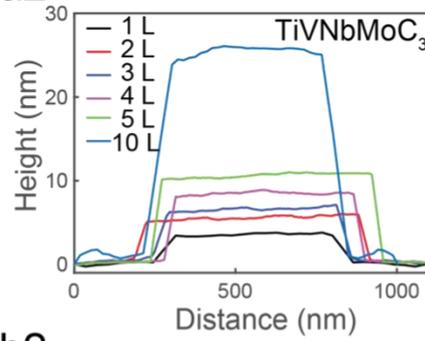 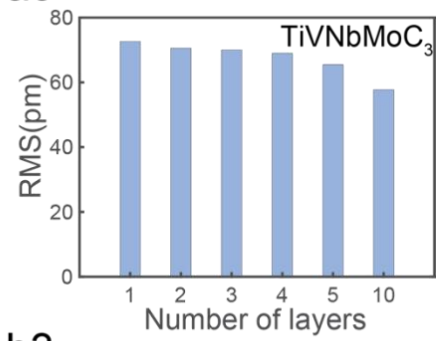
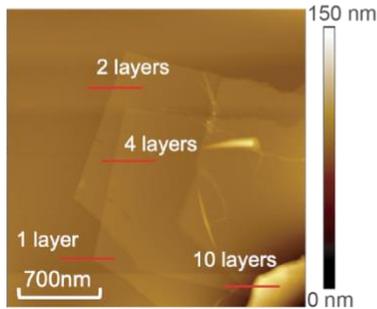 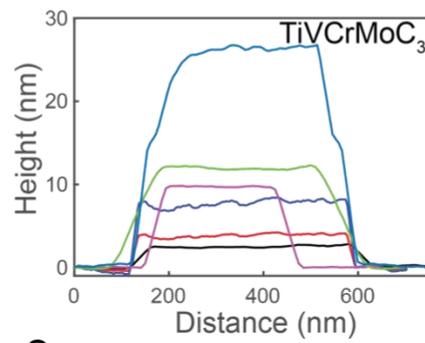 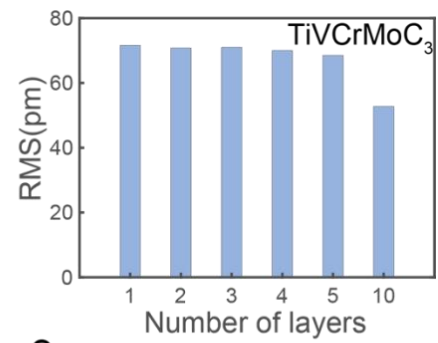
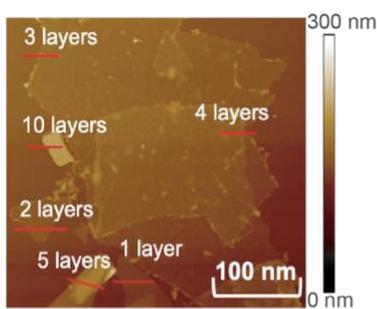 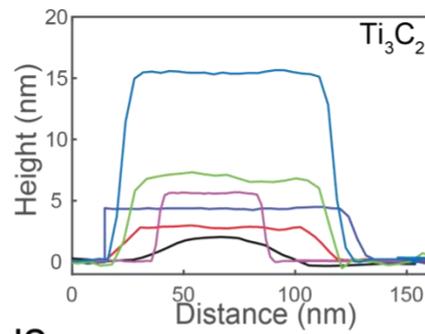 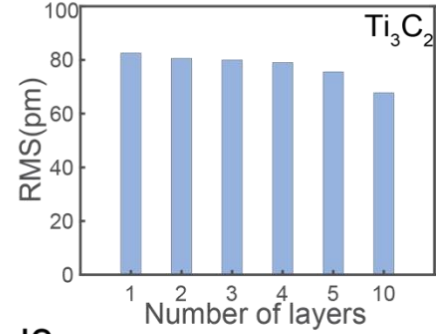
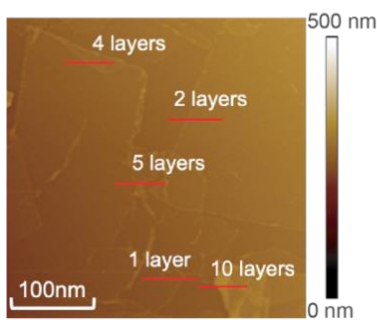 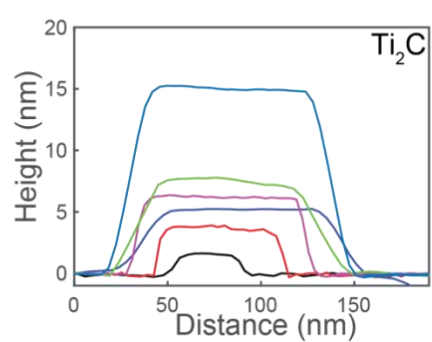 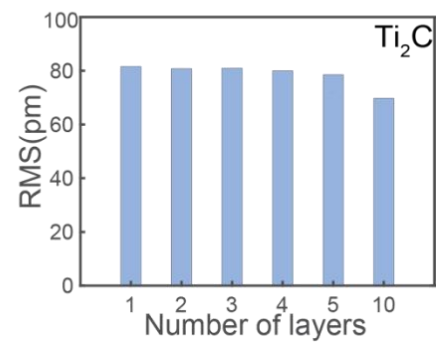

**Figure 2. Topography and morphological analysis of MXenes.** AFM scanning of **a** TiVNbMoC$_3$, **b** TiVCrMoC$_3$, **c** Ti$_2$C, and **d** Ti$_3$C$_2$ MXene. For each MXene, the panels from top to bottom display the AFM image (1), surface height (2), and surface roughness (3) across different layers. Note: To ensure sample freshness, all tests on MXenes were conducted within 36 hours of intercalation. The layer thickness and surface roughness for all four types of MXenes were measured over the same length scale to ensure consistency and fair comparison.

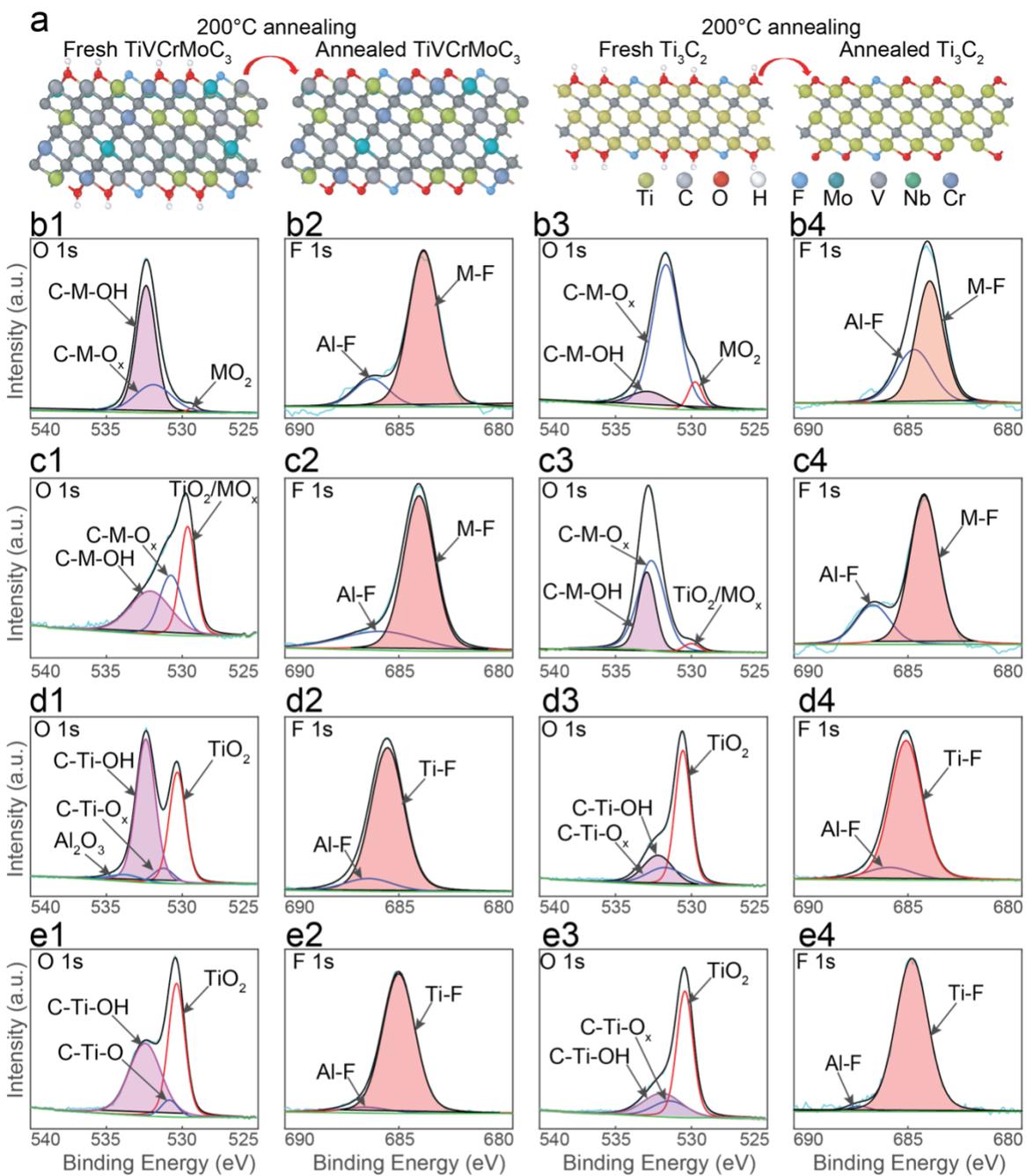

**Figure 3. Surface chemistry and termination analysis of MXenes. a** Schematic illustrating the atomic structural changes in ME MXenes and Ti-C MXenes before and after thermal annealing; High-resolution XPS detection on the terminating group of MXenes: **b** TiVNbMoC$_3$, **c** TiVCrMoC$_3$, **d** Ti$_2$C, and **e** Ti$_3$C$_2$. For each MXene, the panels from left to right show O 1s and F 1s, at fresh (1)(2) and annealed condition (3)(4), respectively; **f** Relative ratios of terminating groups for MXenes under fresh and annealed conditions. The dashed line in the figure represents the average relative ratio of -OH groups (blue) and -O groups (red). An asterisk (*) indicates annealed MXenes.

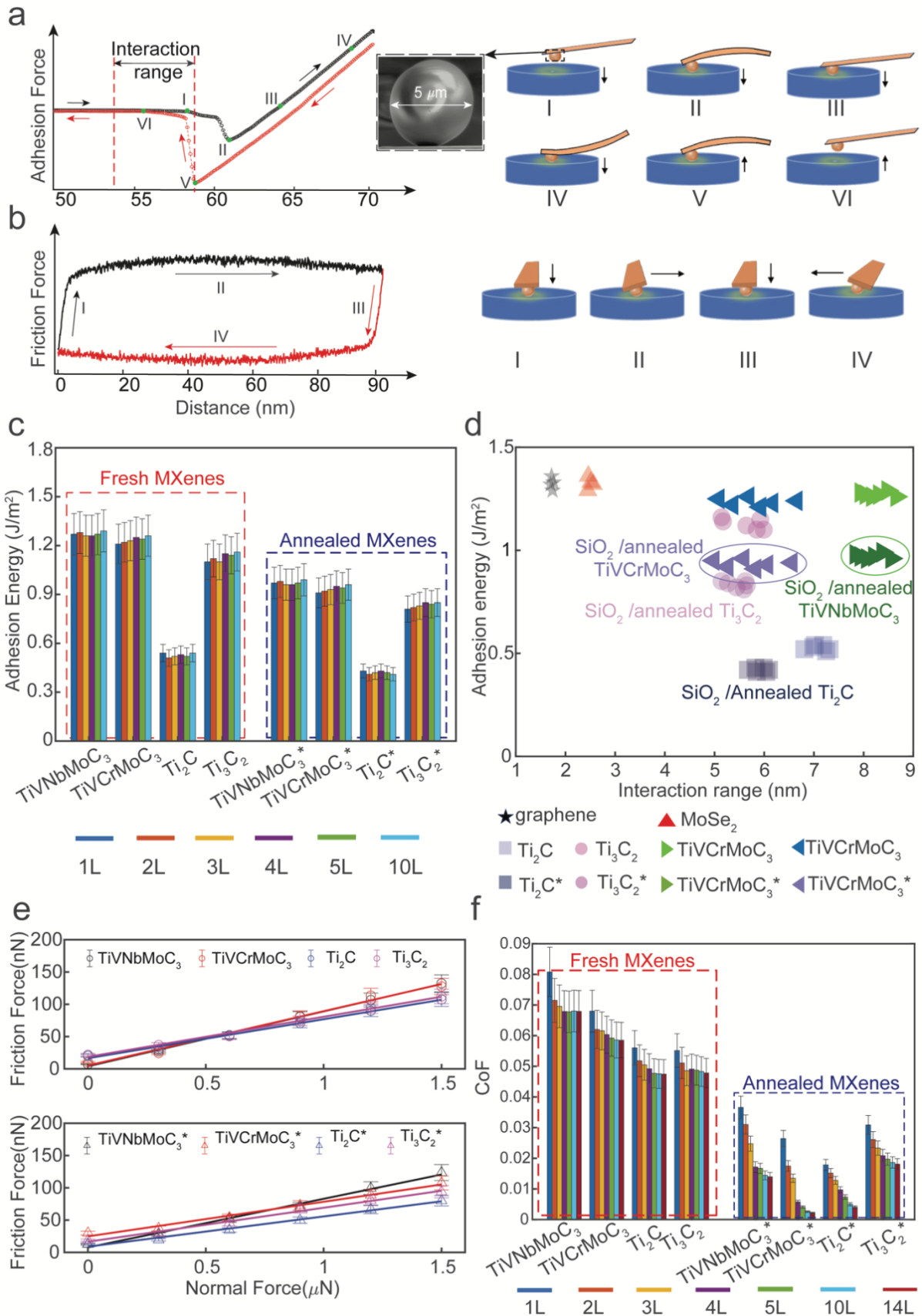

**Figure 4. Adhesion and friction analysis of MXenes**. Schematic of the mechanical measurements: **a** adhesion measurement and **b** friction measurement; Analysis results of adhesion performance: **c** adhesion energy of fresh and annealed MXenes, and **d** adhesion energy *versus* interaction range for MXenes and other 2D materials; Analysis results of friction performance: **e** friction force of 14-layers MXenes under fresh and annealed conditions; **f** CoF of MXenes under fresh and annealed conditions. An asterisk (*) indicates annealed MXenes.

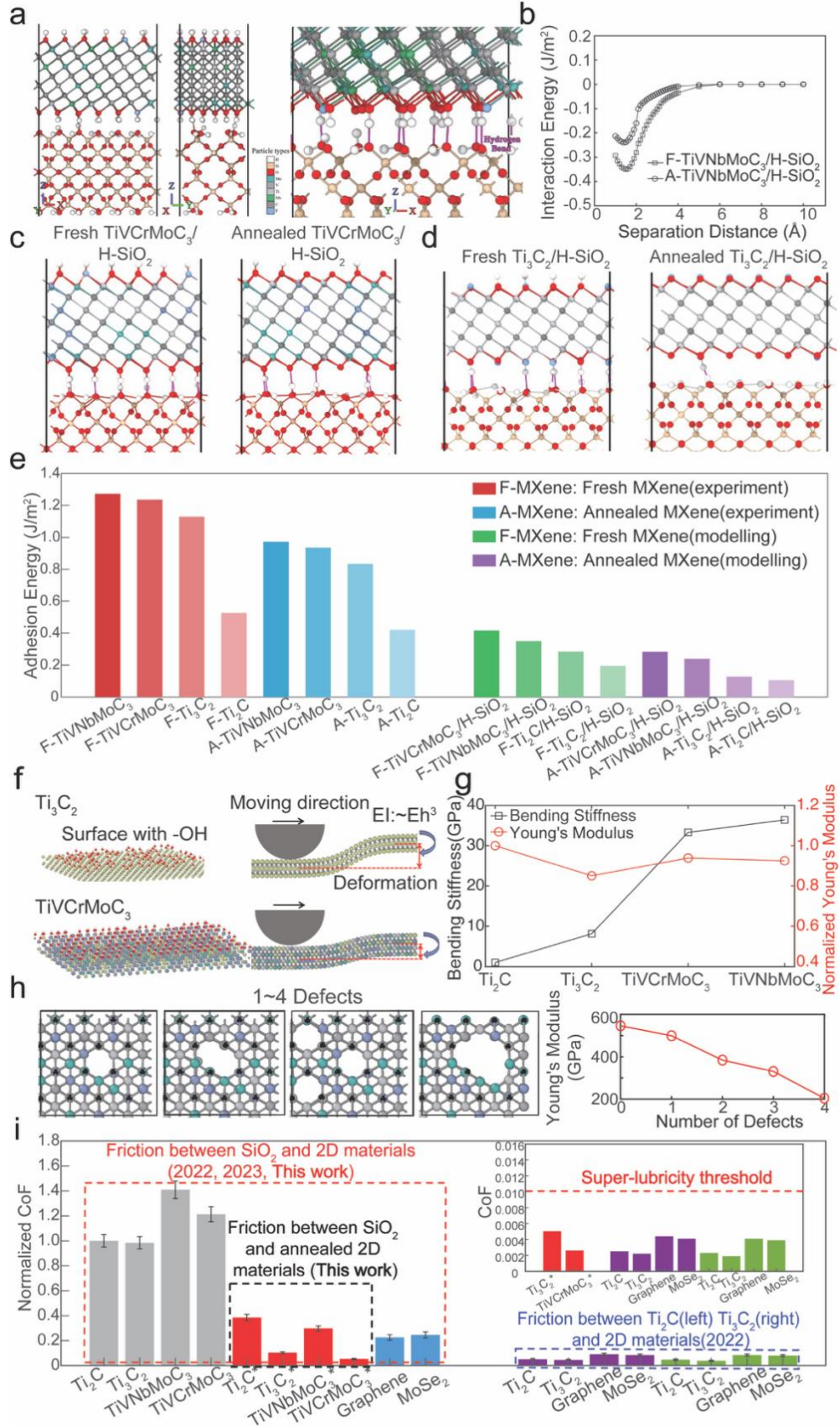

**Figure 5. DFT and MD Calculations on interfacial adhesion and mechanical properties of MXenes. a** Representative minimized TiVNbMoC$_3$/hydroxylated SiO$_2$ supercell at the optimal separation (cell dimensions ~ 0.9×1.5×3.0 nm³). And interfacial hydrogen bonds at the TiVNbMoC$_3$/ hydroxylated SiO$_2$ interface, highlighted in pink. **b** Interaction-energy profiles for fresh and annealed TiVNbMoC$_3$ on hydroxylated SiO$_2$. **c, d** Atomic schematics of hydrogen bonding at fresh (left) and annealed (right) MXene/hydroxylated SiO$_2$ interfaces: **c** TiVNbMoC$_3$ and **d** Ti$_3$C$_2$. **e** Adhesion energy of MXenes determined from experiments and atomistic simulations. **f** Schematic representation of surface terminations and bending stiffness for ME MXenes and Ti-C MXenes. **g** Bending stiffness and normalized Young's modulus of MXenes. **h** Atomic simulations of surface defects with varying defect densities on the ME MXene surface. The plot highlights defect-induced variations in the Young's modulus of ME MXenes. **i** Friction performance landscape of various 2D materials. The inset shows the specific 2D materials exhibiting superlubricity. An asterisk (*) indicates annealed materials.

# Reference

1. Wyatt, B. C., Yang, Y., Michałowski, P. P., Parker, T., Morency, Y., Urban, F. & Anasori, B. Order-to-disorder transition due to entropy in layered and 2D carbides. *Science* **389**, 1054-1058 (2025).
2. Leong, Z., Jin, H., Wong, Z. M., Nemani, K., Anasori, B., & Tan, T. L. Elucidating the chemical order and disorder in high-entropy MXenes: A high-throughput survey of the atomic configurations in $TiVNbMoC_3$ and $TiVCrMoC_3$. *Chem Mater*. **34**, 9062-9071 (2022).
3. Ding, Q., et al. Tuning element distribution, structure and properties by composition in high-entropy alloys. *Nat*. **574**, 223-227 (2019).
4. Zhou, C., Li, Z., Liu, S., Ma, L., Zhan, T., & Wang, J. Synthesis of MXene-based self-dispersing additives for enhanced tribological properties. *Tribol. Lett*. **70**, 63 (2022).
5. Du, Z., et al. High-entropy atomic layers of transition-metal carbides (MXenes). *Adv. Mater*. **33**, 2101473 (2021).
6. Firouzjaei, M. D., Karimiziarani, M., Moradkhani, H., Elliott, M., & Anasori, B. MXenes: The two-dimensional influencers. *Mater. Today Adv*. **13**, 100202 (2022).
7. Du, C. F., et al. Synthesis of a high-entropy $(TiVCrMo)_3AlC_2$ MAX and its tribological properties in a wide temperature range. *J. Eur. Ceram*. **43**, 4684-4695 (2023).
8. Wyatt, B. C., Rosenkranz, A., & Anasori, B. 2D MXenes: tunable mechanical and tribological properties. *Adv. Mater*. **33**, 2007973 (2021).
9. Caffrey, N. M. Effect of mixed surface terminations on the structural and electrochemical properties of two-dimensional $Ti_3C_2T_2$ and $V_2CT_2$ MXenes multilayers. *Nanoscale*. **10**, 13520-13530 (2018).
10. Lee, C., Li, Q., Kalb, W., Liu, X.Z., Berger, H., Carpick, R.W. & Hone, J. Frictional characteristics of atomically thin sheets. *Science.* **328**, 76-80 (2010).
11. Tripathi, M. *et al.* Friction and adhesion of different structural defects of graphene. *ACS Appl. Mater. Interfaces* **10**, 44614–44623 (2018).
12. Li, Y., et al. Adhesion of two-dimensional titanium carbides (MXenes) and graphene to silicon. *Nat. Commun*. **10**, 3014 (2019).
13. Li, Y., et al. Adhesion between MXenes and other 2D materials. *ACS Appl. Mater. Interfaces*. **13**, 4682-4691 (2021).
14. Li, Y., et al. Friction between MXenes and other two-dimensional materials at the nanoscale. *Carbon*. **196**, 774-782 (2022).
15. Machata, P., et al. Wettability of MXene films. *J. Colloid Interface Sci*. 622, 759-768 (2022).
16. Magnuson, M., Halim, J., & Näslund, L. Å. Chemical bonding in carbide MXene nanosheets. *J. Electron Spectrosc. Relat. Phenomena*. **224**, 27-32 (2018).
17. Guan, Y., Zhang, M., Qin, J., Ma, X., Li, C., & Tang, J. Hydrophilicity-dependent distinct frictional behaviors of different modified MXene nanosheets. *J. Phys. Chem. C*. **124**, 13664-13671 (2020).

18. Hu, T., et al. Interlayer coupling in two-dimensional titanium carbide MXenes. *PCCP*. **18**, 20256-20260 (2016).
19. Rosenkranz, A., Righi, M. C., Sumant, A. V., Anasori, B., & Mochalin, V. N. Perspectives of 2D MXene tribology. *Adv. Mater*. **35**, 2207757 (2023).
20. Yu, H., et al. Mapping the structure and chemical composition of MAX phase ceramics for their high-temperature tribological behaviors. *CE*. e597 (2024).
21. Serles, P., et al. Friction of $Ti_3C_2T_x$ MXenes. *Nano Lett*. **22**, 3356-3363 (2022).
22. Guo, Y., Zhou, X., Wang, D., Xu, X., & Xu, Q. Nanomechanical properties of $Ti_3C_2$ MXene. *Langmuir*. **35**, 14481-14485 (2019).
23. Zhang, D., et al. Computational study of low interlayer friction in $Ti_{n+1}C_n$ (n= 1, 2, and 3) MXene. *ACS Appl. Mater. Interfaces*. **9**, 34467-34479 (2017).
24. Zhai, W., et al. Recent progress on wear-resistant materials: designs, properties, and applications. *Adv. Sci*. **8**, 2003739. (2021).
25. Stoyanov, P., & Chromik, R. R. Scaling effects on materials tribology: from macro to micro scale. *Materials*. **10**, 550 (2017).
26. Pandey, R. P., et al. Effect of sheet size and atomic structure on the antibacterial activity of Nb-MXene nanosheets. *ACS Appl. Nano Mater*. **3**, 11372-11382 (2020).
27. Li, Q., Lee, C., Carpick, R. W. & Hone, J. Substrate effect on thickness-dependent friction on graphene. *Phys. Status Solidi B* **247**, 2909–2914 (2010).
28. Anasori, B. & Gogotsi, Y. MXenes: trends, growth, and future directions. *Graphene 2D Mater*. **7**, 75–79 (2022).
29. Madsen, J., Susi, T. The abTEM code: transmission electron microscopy from first principles. *Open Research Europe*. **1**, 24 (2021).
30. Mehtab, S., Zaidi, M. G. H., & Siddiqi, T. I. Designing fructose stabilized silver nanoparticles for mercury (II) detection and potential antibacterial agents. *Mater. Sci. Res. India*. **15**, 241-249 (2018).
31. Chen, T., et al. Accelerating hole extraction by inserting 2D $Ti_3C_2$-MXene interlayer to all inorganic perovskite solar cells with long-term stability. *J Mater Chem A*. **7**, 20597-20603 (2019).
32. Lin, Z., et al. High-yield exfoliation of 2D semiconductor monolayers and reassembly of organic/inorganic artificial superlattices. *Chem*. **7**, 1887-1902 (2021).
33. Nemani, S. K., et al. High-entropy 2D carbide MXenes: $TiVNbMoC_3$ and $TiVCrMoC_3$. *ACS Nano*. **15**, 12815-12825 (2021).
34. Kang, R., et al. Enhanced thermal conductivity of epoxy composites filled with 2D transition metal carbides (MXenes) with ultralow loading. *Sci. Rep*. **9**, 9135 (2019).
35. Lu, Y., Li, D., & Liu, F. Characterizing the chemical structure of $Ti_3C_2T_x$ MXene by angle-resolved XPS combined with argon ion etching. *Materials*. **15**, 307 (2022).
36. Jacobs, T. D. B. et al. The effect of atomic-scale roughness on the adhesion of nanoscale asperities: a combined simulation and experimental investigation. *Tribol. Lett*. **50**, 81-93 (2013).
37. Rabinovich, Y. I., Adler, J. J., Ata, A., Singh, R. K. & Moudgil, B. M. Adhesion between nanoscale rough surfaces: I. role of asperity geometry. *J. Colloid Interface Sci*. **232**, 10-16 (2000).

38. Marquis, E., Cutini, M., Anasori, B., Rosenkranz, A. & Righi, M. C. Nanoscale MXene interlayer and substrate adhesion for lubrication: a density functional theory study. *ACS Appl. Nano Mater.* **5**, 10516-10527 (2022).
39. Rosenkranz, A. *et al.* Multi-layer $Ti_3C_2T_x$ nanoparticles (MXenes) as solid lubricants-role of surface terminations and intercalated water. *Appl. Surf. Sci.* **494**, 13-21 (2019).
40. Arif, T., Colas, G. & Filleter, T. Effect of humidity and water intercalation on the tribological behavior of graphene and graphene oxide. *ACS Appl. Mater. Interfaces* **10**, 22537-22544 (2018).
41. Arif, T., Wang, G., Sodhi, R. N. S., Colas, G. & Filleter, T. Role of chemical vs. physical interfacial interaction and adsorbed water on the tribology of ultrathin 2D-material/steel interfaces. *Tribol. Int.* **163**, 107194 (2021).
42. Zaretsky, E. V. Bearing and gear steels for aerospace applications. *NASA Technical Memorandum* 102026 (1990).
43. Przybyła, R. *et al.* Improvement of tribological properties of 440C steel in terms of applications in space mechanisms. *Tribol. Trans.* **58**, 832-840 (2015).
44. Janna, William S. Engineering Heat Transfer. *CRC press*. Boca Raton, (2018)
45. Deng, Z., Smolyanitsky, A., Li, Q., Feng, X. Q., & Cannara, R. J. Adhesion-dependent negative friction coefficient on chemically modified graphite at the nanoscale. *Nat. Mater.* **11**, 1032-1037 (2012).
46. Long, F., Yasaei, P., Yao, W., Salehi-Khojin, A., & Shahbazian-Yassar, R. Anisotropic friction of wrinkled graphene grown by chemical vapor deposition. *ACS Appl. Mater. Interfaces*. **9**, 20922-20927 (2017).

## Acknowledgments

This work was supported by the National Science Foundation #1930881 and #2414716.

## Author contributions

Conceptualization: C.Wu.; Design: C.Wu., R.W.; Experiment: J.L., Y.Z., Y.L., S.H.; Modeling: C.W. R.N., W.G.; Writing-original draft: J.L., C.W., R.N., W.G., C. Wu; Proofreading and editing: J.L., C.W., T.F., B.W., K.X., B.A., A.E., W.G., C.Wu.; Writing-review and editing: all authors; Funding acquisition: C.Wu.

## Competing interests

The authors declare no competing financial interest.

## Additional information

Supplementary materials are available. The data supporting the findings of this study are available from the corresponding author upon reasonable request.